\documentstyle[11pt,epsfig]{article} 
\title{A Comparison of Cosmic Ray Composition Measurements at the Highest Energies}
\author{B.R.\,Dawson, R.\,Meyhandan and K.M.\,Simpson\\Department of Physics and Mathematical Physics\\The University of Adelaide, Australia 5005\\email: bdawson@physics.adelaide.edu.au}
\date{}
\begin{document}
\begin{titlepage}
\maketitle
\begin{abstract}
In recent years the Fly's Eye and Akeno groups have presented
analyses of the cosmic ray mass composition at energies above
$10^{17}$eV.  While the analysis of the Fly's Eye group points to
a likely change in mass composition from heavy to light at
energies above $10^{18}$eV, the Akeno analysis favours an
unchanging composition.  However, the two groups base their
conclusions on simulations using quite different hadronic models.
Here we present a comparison of the experiments using the same
hadronic model and find that the agreement between the
experiments is much improved.  Under this model, both experiments
measure a composition rich in iron around $10^{17}$eV which
becomes lighter at higher energies.  However, the agreement is
not complete, which indicates scope for improvement of the
interaction model, or perhaps the need for a re-examination of the
experimental results.
\end{abstract}
\centerline{{\em To appear in Astroparticle Physics}}
\centerline{astro-ph/9801260}
\centerline{{\em Revised 21 May 1998}}
\end{titlepage}

\section{Introduction}
Measurement of the mass composition of cosmic rays at the highest
energies is a challenging and vital task.  A known mass (and
therefore charge) spectrum can provide strong constraints on the
acceleration and propagation of the highest energy particles
known in the Universe.  Such a measurement is as important as
measurements of the energy spectrum or anisotropy.

Mass composition measurements are difficult to interpret because
of quite understandable deficiencies in our knowledge of hadronic
interactions at the highest energies.  This is particularly true
at energies discussed here, with primary cosmic ray energies
above $10^{17}$eV.  However, progress is being made in these
models, with some constraints being imposed by cosmic ray data.
Our intention here is to show that two sets of mass composition
data, previously thought to be inconsistent, are in fact somewhat
reconcilable.  The degree to which the experimental results agree
could be a test of the validity of a given interaction model.

Two very different experiments have reported mass
composition interpretations in the recent past.  The University
of Utah Fly's Eye detector \cite{Balt85} observed the
longitudinal development of energetic air showers via the
emission of fluorescence light in the atmosphere.  This technique
effectively uses the atmosphere as a calorimeter and extracts
both energy and mass composition information from the
longitudinal profile.  Heavy nuclei produce showers which
develop, on average, more rapidly than proton showers of the same
energy, and the depth at which a shower reaches its maximum size,
$X_{\rm max}$ is used as the mass composition indicator.  

A complementary technique is employed by the Akeno group in
Japan. Their giant array AGASA \cite{AGASA} covers 100\,km$^2$
and contains shielded and unshielded detectors to measure,
respectively, the muon and charged particle densities at the
detector depth of 920 g\,cm$^{-2}$.  Here the energy is estimated
by $S_0(600)$, the unshielded scintillator density 600\,m from
the shower core corrected for zenith angle.  The mass composition
is characterised by the muon measurements, with an expectation
that showers initiated by heavy nuclei will, on average, have a
higher muon content at a particular energy.  

As already mentioned, the interpretation of the results from
these experiments requires assumptions about hadronic
interactions at energies well beyond the reach of accelerators.
In \cite{Gaiss93} the Fly's Eye group investigates 3 different
interaction models, one of which is ruled out by the $X_{\rm
max}$ data.  The two remaining models, known as KNP and minijet,
are both characterised by increases in inelasticity with energy
(inelasticity being the fraction of energy used in an interaction
in the production of secondary particles), with the KNP model
said to provide close to an upper bound on the rate of increase
in inelasticity \cite{Todor}.  These models predict rather low
values of the elongation rate ($d\log(X_{\rm max})/d\log E$)
which are quite inconsistent with the data.  The conclusion from
both models is that the data indicate a decrease in the mean
cosmic ray mass with energy, with an iron dominated flux around
$10^{17}$ eV and a significantly lighter one at around $10^{19}$
eV.

The Akeno group base their composition conclusions \cite{Haya95}
on the hadronic model contained in the 1992 version of MOCCA, an
air shower simulation written by A.M.\,Hillas.  This hadronic
model is based on results from fixed target accelerator
experiments.  The model makes a prediction for the way the muon
content (as expressed by the density of muons 600\,m from the
shower core) increases with energy.  Based on this expectation,
the Akeno group see no evidence for a change in composition in
the same energy range explored by the Fly's Eye experiment.

In recent years a new hadronic interaction generator named SIBYLL
has been used within the MOCCA shower simulation code.  We have used
this generator to investigate the longitudinal development and
muon content of large air showers for comparison with the
experimental data.

\section{The SIBYLL Hadronic Generator}
The SIBYLL interaction generator \cite{Flet94} is built around
the ``minijet'' model used in the Fly's Eye analysis, and was
designed to reproduce features observed in collider experiments.
As we have said, the original MOCCA hadronic driver (as used by
the Akeno group \cite{Haya95}) was written to reproduce
interactions at more modest fixed target accelerator energies.
The differences in physics are described in \cite{Flet94}, but
briefly the original MOCCA featured a flat Feynman-$x$
distribution of the leading nucleon, Feynman scaling of resulting
momentum distributions, and charged particle multiplicity
distributions with mean values increasing as $\ln(s)$ ($s$ being
the square of the centre of mass energy).  In contrast, SIBYLL
uses a model which has multiplicities increasing more rapidly
with energy (as $\ln^2s$), and the minijet character of the model
allows a more sophisticated treatment of correlations between
transverse momentum and multiplicity.  Scaling is violated, but
not in an extreme way.  The authors regard SIBYLL as a
lower-bound on scaling violations.  Compared with the model
within the original MOCCA, SIBYLL predicts air showers with a
more rapid development, with expected differences in measurable
quantities like the elongation rate and muon content.

We have generated vertical and inclined proton and iron showers
at several energies using MOCCA+SIBYLL at a thinning level of
$10^{-6}$ (see \cite{Hill81} for a description of thinning).  We
record the longitudinal development profiles, and ground particle
information appropriate for the AGASA atmospheric depth.

\section{Data Sets}
We compare simulation results with data from the Fly's Eye and
Akeno experiments.  The Fly's Eye data set is that published in
\cite{Gaiss93}.  In the case of Akeno, we have used different
sources for the data collected by the A1 and A100 arrays.  The
full A1 array \cite{A1} began operation in October 1981, covers
an area of approximately 1\,km$^2$, and is populated with surface
scintillator detectors and shielded muon detectors.  The latter
have a muon energy threshold for vertical particles of 1\,GeV.
Our analysis of this array is based on data from \cite{Haya95}.

The A100, or AGASA, array covers an area of 100\,km$^2$ and has
operated since 1992.  Roughly half of this area is also serviced
by muon detectors, in this case with a vertical muon energy
threshold of 0.5\,GeV.  We have used a recent set of data from
this array \cite{Haya97}.  This represents more than a seven-fold
increase in the size of the data set presented for A100 in the
original analysis \cite{Haya95}.

In both papers, the practice of the authors has been to reduce
the A100 muon densities $\rho_\mu(600)$ by a factor of 1.4, to
allow comparison with A1 data (A1 having a higher muon energy
threshold) and their simulations, which were calculated for a
threshold of 1\,GeV.  We prefer to separately compare A1 data
with 1\,GeV simulations and A100 data with 0.5\,GeV simulations,
since we have seen in the simulations that the 1.4 factor, though
experimentally measured at one energy ($10^{17}$eV \cite{1.4}),
appears not to be valid for all energies and mass compositions.
For example, our MOCCA+SIBYLL simulations show a ratio of $1.41\pm0.01$
for iron at $10^{17}$eV, but a value of $1.19\pm0.01$ for protons at the
same energy.


In \cite{Haya97} the A100 muon densities are given as a function
of primary energy.  In \cite{Haya95} A1 muon densities are given
as a function of the scintillator density 600\,m from the shower
core, $S(600)$.  In each case the sample of showers has a 
$<\sec\theta>=1.09$ ($\theta$ being the zenith angle).  Later we
will plot these A1 data as a function of primary energy.  The
primary energy has been calculated from $S(600)$ in two steps.
First $S_0(600)$, the expected scintillator density for vertical
showers is calculated, assuming a shower attenuation length of
500\,g\,cm$^{-2}$ \cite{attenn}.  The primary energy is then
assumed to be $E=2\times10^{17}S_0(600)$ \cite{energy,dai88}.

\section{Treatment of Detector Effects}
It is important to understand experimental biases in the analysis
of composition data.  These biases can occur in the triggering of
the experiment or in the analysis of the data.  The triggering
and reconstruction biases of the Fly's Eye have previously been
studied in great detail by ourselves and others.  We have used
one such study \cite{Ding97} to apply corrections to the
MOCCA+SIBYLL-simulated elongation rate for proton and iron
showers.  That study generated proton and iron showers with
energies sampled from a $E^{-3}$ differential spectrum.  The
showers were passed through a detailed detector simulation and
reconstructed using the same analysis routines used for the real
data.  Such a process takes account of effects like triggering
bias (there is a small bias in the Fly's Eye against detecting
iron showers around the threshold energy of $10^{17}$eV), and
small shifts in reconstructed values of $X_{\rm max}$, this
partly due to the assumption of a gaussian form for the
longitudinal profile.

The $X_{\rm max}$ behaviour of our MOCCA+SIBYLL generated showers
is similar to that used in \cite{Ding97}.  At a given energy,
proton $X_{\rm max}$ values differ by around 20 g\,cm$^{-2}$ and
iron values differ by around 30 g\,cm$^{-2}$.  From \cite{Ding97}
we parametrize triggering and reconstruction shifts in terms of
energy and $X_{\rm max}$.  Because of the two-dimensional
parametrization we feel confident applying these shifts to our
data, despite slight differences in $X_{\rm max}$.  In addition
to these shifts we have reduced all simulated $X_{\rm max}$
values by 20 g\,cm$^{-2}$.  This was done in the original study
\cite{Gaiss93}, and is also needed here to ensure that the real
data at $10^{17}$eV do not imply a composition {\em heavier} than
iron.  Such a shift is entirely consistent with possible systematic
effects in the Fly's Eye experiment \cite{Gaiss93}.

To investigate any trigger bias in the Akeno experiments, we take
showers from MOCCA+SIBYLL and simulate their interaction with the
A1 and A100 arrays.  For each simulated shower we obtain the muon
lateral distribution and the lateral distribution expected for
5\,cm thick plastic scintillator detectors.  The muon energy
threshold is assumed to be 1.0\,GeV for the A1 array and 0.5\,GeV
for the A100 array.  The showers are thrown at random core
locations within the arrays, the densities are statistically
fluctuated, and the trigger conditions are applied for both the
scintillator detectors and the muon detectors.  

We find that at energies of interest above $10^{17}$eV the
densely instrumented A1 array is free of triggering bias, with
equal triggering efficiencies for both proton and iron-induced
EAS in both the scintillator and muon detector parts of the
array.  We assume that this array will, on average, correctly
measure the two shower parameters of interest - the scintillator
density $S(600)$ and the muon density $\rho_\mu(600)$ at a core
distance of 600\,m.  

In contrast we find that the larger and sparser A100 array has
some composition-dependent triggering efficiency.  To investigate
this, we generated simulated showers at a zenith angle of 23
degrees, to match the mean $\sec\theta$=1.09 of the data set.
(The A1 simulations were performed with vertical showers.  We
determined that an extra 9\% of atmospheric depth would not
affect our conclusion regarding any possible triggering bias).

The Akeno group has opted for an A100 triggering scheme which
minimizes trigger bias.  We have applied this scheme in our
simulations.  The first requirement for a shower is that it
triggers the surface array, with a minimum of 5 nearest-neighbour
stations registering particles.  A muon station will self-trigger
if two (non-neighbouring) proportional counters fire within that
station.  A muon station may also be triggered (and data read
out) if it fails to meet the 2 proportional counter requirement,
but only if the scintillator detector at the same location
records a signal above threshold.  This is known as an assisted
muon trigger.  A shower is accepted for the analysis if it
satisfies the electron array trigger and if there is at least one
self-triggered or assisted-triggered muon station.  Obviously,
some of the assisted muon triggers record a zero particle count.

If the self- or assisted-triggered muon station is at a core
distance between 500 and 800\,m it is used in the calculation of
an average $\rho_\mu(600)$ for the appropriate $S(600)$ bin.
Every such muon density is scaled to the expected density at
600\,m using the published AGASA muon lateral distribution
function \cite{Haya95}.  The resulting scaled densities are
averaged to find the mean $\rho_\mu(600)$.  Note that some
showers might contribute 5 or more numbers to the average.  Some
other showers contribute a single zero to the average.  The
latter case occurs when the only triggered muon station in that
event is an assisted trigger with a measured density of zero.
(This seems a good strategy.  A measured muon count of zero
is a perfectly valid fluctuation from a mean expectation that might be as
low as 0.2 muon m$^{-2}$ at the lowest energies).

Given the A100 triggering scheme, we simulated the response of
the array to our MOCCA-generated showers, and calculated the mean
$\rho_\mu(600)$ for those showers triggering at each energy.  We
also took note of the triggering efficiency as a function of
energy for proton and iron initiated showers.  We find that at
$3\times10^{17}$eV the triggering efficiency (electron+muon) for
proton showers is 80\% of the iron shower efficiency.  The
efficiencies equalize above $3\times10^{18}$eV.  These figures are
used later in our estimate of the fraction of iron present in the
primary cosmic ray beam.

In graphs that follow we will plot the energies of MOCCA
simulations.  We do this in the following way.  In the case of
the vertical showers from the A1 simulation, we take from the
simulations the mean value of the scintillator density at 600\,m,
S(600).  (Of course, this is actually $S_0(600)$ because the
showers are vertical).  We convert this to primary energy
assuming the Akeno conversion factor $E=2\times10^{17}S_0(600)$
\cite{energy,dai88}.  In the case of the A100 simulation with
inclined showers, we follow the same procedure, except that we
convert the measured mean S(600) into $S_0(600)$ by assuming a
shower attenuation length of 500\,g\,cm$^{-2}$.  While the
resulting energies in both the A1 and A100 simulations appear
somewhat inconsistent with the primary energies injected into the
original MOCCA+SIBYLL simulation, we do this to follow the
procedure used with the real data.

\section{Results and Discussion}
Our comparison of Fly's Eye $X_{\rm max}$ data with MOCCA+SIBYLL
(Figure~1) is similar to the result obtained by previous authors
using the minijet model \cite{Gaiss93}.  The simulated elongation
rates for proton and iron showers are significantly smaller than
the elongation rate observed by the Fly's Eye, and so we conclude
that the model requires a change from an iron-dominated
composition to a lighter one above $10^{18}$eV.  The change in
composition is not as striking as that required by the more
extreme KNP model \cite{Gaiss93}, but a change is nonetheless
required.

We present results from our Akeno analysis in Figures~2\,\&\,3.
Here we show data from A1 and A100 compared with simulations
appropriate to the muon energy threshold for the particular
array.  

A note on the energy scales in these two figures: as described
above, the simulation points (see figures) are assigned energies
on the basis of the mean value of $S_0(600)$ and the Akeno energy
conversion factor.  This results in the assigned energies being
low compared to the primary energies injected into the
simulation. That is, according to the SIBYLL model, the Akeno
procedure {\em underestimates} primary energy by around
30\%. (Actually, the apparent underestimation is around 30\% for
the vertical showers used in the A1 simulation and 20\% for
inclined showers used in the A100 simulation. There is no obvious
energy dependence to these figures).  Such an inconsistency was
noted in the original Akeno paper on the $S(600)$ to energy
conversion \cite{dai88} in which the conversion factor was
compared with other experiments and models.  Not surprisingly,
there is model dependence in the conversion factor.

In Figure~4 we show the implied fraction of iron nuclei in the
primary beam at the top of the atmosphere, from A1 and A100 data
assuming a simple mixture of protons and iron nuclei.  For A100
data we have included the effect of the small trigger bias in
favour of iron showers at the lowest energies.  The line on the figure
represents a fit to the A1 data.  We note that there is good agreement
between the A1 and A100 results, encouraging since they are separate 
experiments with different muon energy thresholds and analysis methods.

In Figure~5 we show the corresponding plot for the Fly's Eye analysis.
We include in this figure the A1 array best-fit line from Figure~4.
Comparing the results, we observe the following

\begin{itemize}

\item  Both AGASA and Fly's Eye support a heavy composition (100\% iron
in the two component model) near $10^{17}$eV.

\item Both experiments require that the composition becomes
significantly lighter in the energy range up to
$3\times10^{18}$eV.  At that energy, both experiments require a
fraction of iron of around 55-70\% in this two component fit
under the SIBYLL model.  Above this energy, statistical
uncertainties preclude a definitive statement.

\item The way in which the fraction of iron changes from 100\% to
around 60\% is different in the two experiments.  The change in the
Akeno data occurs at lower energy.  One possible explanation for
this concerns the energy calibration.  For illustration purposes,
we will assume that AGASA underestimates primary energy by 30\%.
This happens to be the disagreement with the energy scale of
SIBYLL, but we might have chosen to suppose that the Fly's Eye
overestimates energy by the same amount.  This allows us to shift
the A1 array line in Figure~5 giving slightly better agreement
between the experiments.

\item In the spirit of investigating possible experimental
systematics, we have assumed that the systematic error in $X_{\rm
max}$ assignment might be as large as 30\,g\,cm$^{-2}$ rather
than the 20\,g\,cm$^{-2}$ assumed by us above, and as assumed by
Gaisser et al.\cite{Gaiss93}.  Figure~6 shows the fraction of
iron plot for Fly's Eye under this assumption.  This, together
with 30\% shift in the A1 energy scale, brings the results into
better agreement.

\item Of course, it is just as likely that there are no experimental
systematic problems with energy or $X_{\rm max}$.  The difference
between the experimental results may result from physics
assumptions within the SIBYLL model.  There may be scope for
learning something about the model (or its implementation within
the MOCCA shower code) from differences uncovered in this study.

\end{itemize}

Other choices of hadronic model might result in different
conclusions about mass composition.  In particular, the KNP model
used in the original Fly's Eye study required a more rapid change
in composition, from near pure iron to a protonic composition at
the highest energies.  Unfortunately, we did not have access to
this model for 3D calculations of muon density, so it could not
be applied to the AGASA analysis.  It would be interesting to do
such a study to see if less conservative hadronic models give
better agreement between AGASA and Fly's Eye results.

We are aware of a new analysis (in preparation) of EAS seen in
coincidence with the HiRes prototype detector and the MIA muon
array.  While the energy range covered is only from $10^{17}$eV
to $10^{18}$eV, we expect that these results will greatly help
our understanding of correlations between longitudinal shower
development, muon content and mass composition.

\section{Conclusion}
We have used recent data from two experiments that measure the
mass composition of the highest energy cosmic rays in very
different ways.  Under the assumption of a single hadronic model,
we find that the data-sets provide quite similar conclusions.  We
trust this will correct a widely held view that the results are
totally inconsistent.

\section{Acknowledgements}
We greatly appreciate the discussions we have had with members of
the Fly's Eye and Akeno groups, including M.\,Nagano and
Y.\,Matsubara.  We are especially indebted to N.\,Inoue who
provided us with a great deal of insight into the operation of
AGASA, and suffered our many naive questions.

\renewcommand{\topfraction}{1.0}
\renewcommand{\textfraction}{0.0}
\renewcommand{\bottomfraction}{1.0}

\vspace{-1cm}
\begin{figure}[h]
\epsfig{figure=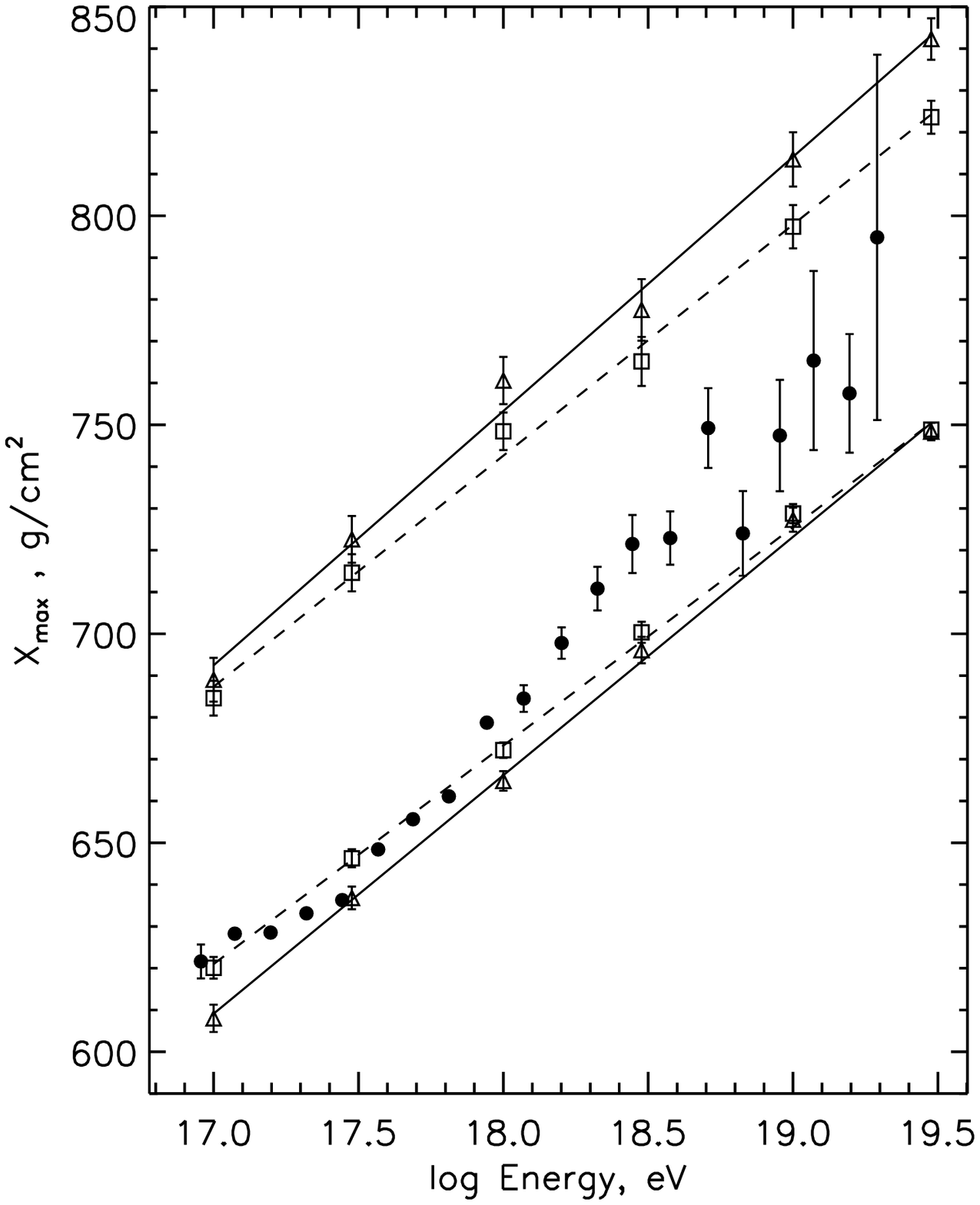,height=18cm} 
\caption{Plot of the Fly's Eye X$_{\rm max}$ vs Energy. 
The solid lines (triangles) are the unreconstructed MOCCA+SIBYLL
simulation data.  The dashed lines (squares) are the
reconstructed simulated data, which also include a coherent 20
g\,cm$^{-2}$ shift (see text). Top lines represent pure protons,
bottom lines represent pure iron.  The uncertainties on the proton
points are larger because of larger intrinsic fluctuations in shower
development.}\
\end{figure}

\begin{figure}[h]
\epsfig{figure=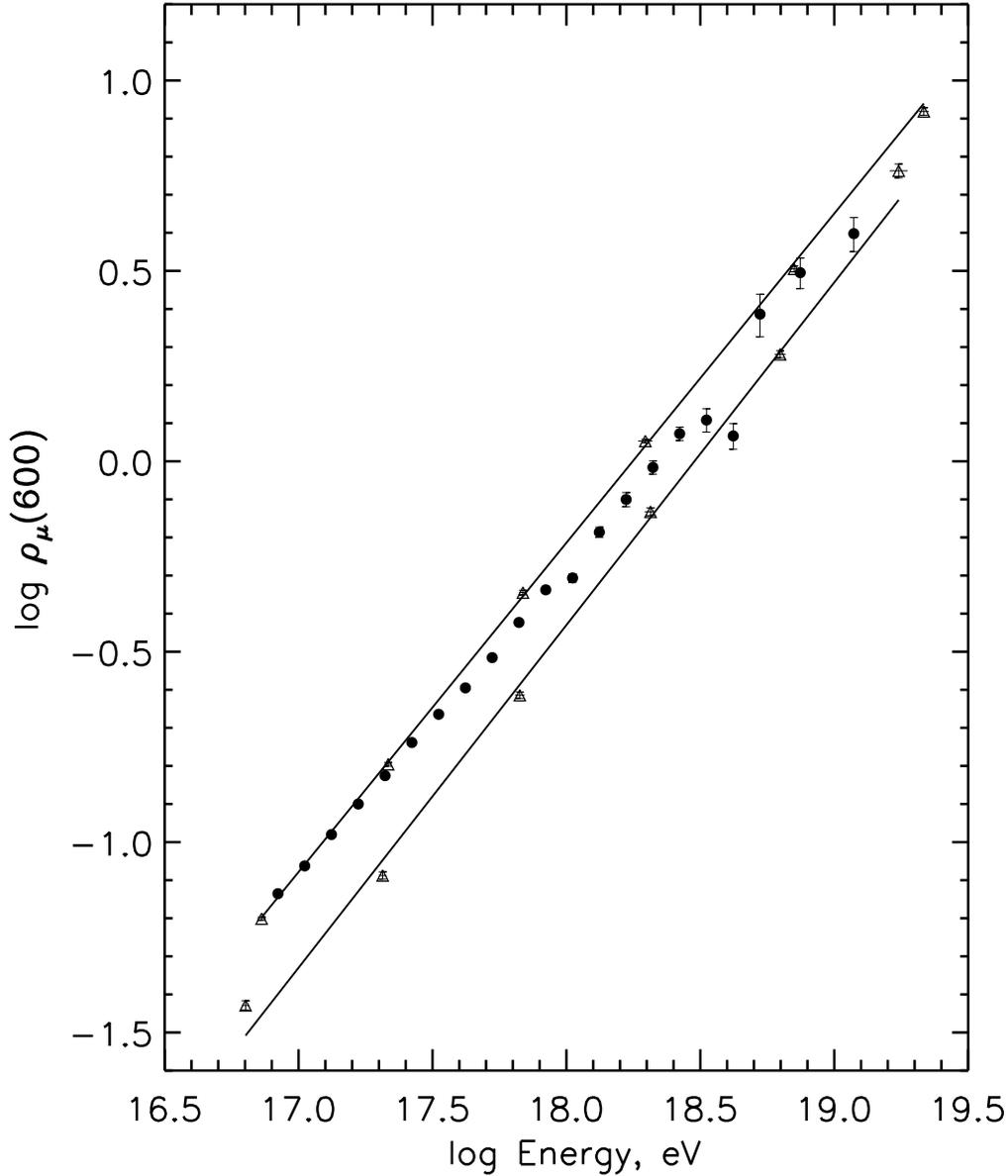,height=17cm}
\caption{Data from the Akeno A1 array up to April 1993 (solid points, \protect\cite{Haya95}) compared with MOCCA+SIBYLL simulations (triangles and solid lines).  The top line represents iron primary particles, the lower line protons.  Calculations were done for a muon energy threshold of 1\,GeV.  Array response simulations show no triggering bias towards either proton or iron-induced showers.
The energy scale is defined using the Akeno method described in
the text.  The scale is somewhat inconsistent with the
MOCCA+SIBYLL calculations - triangles show simulations at
$10^{17}$eV, $3\times10^{17}$eV, $10^{18}$eV, $3\times10^{18}$eV,
$10^{19}$eV and $3\times10^{19}$eV.}
\end{figure}

\begin{figure}[h]
\vspace{-1.4cm}
\epsfig{figure=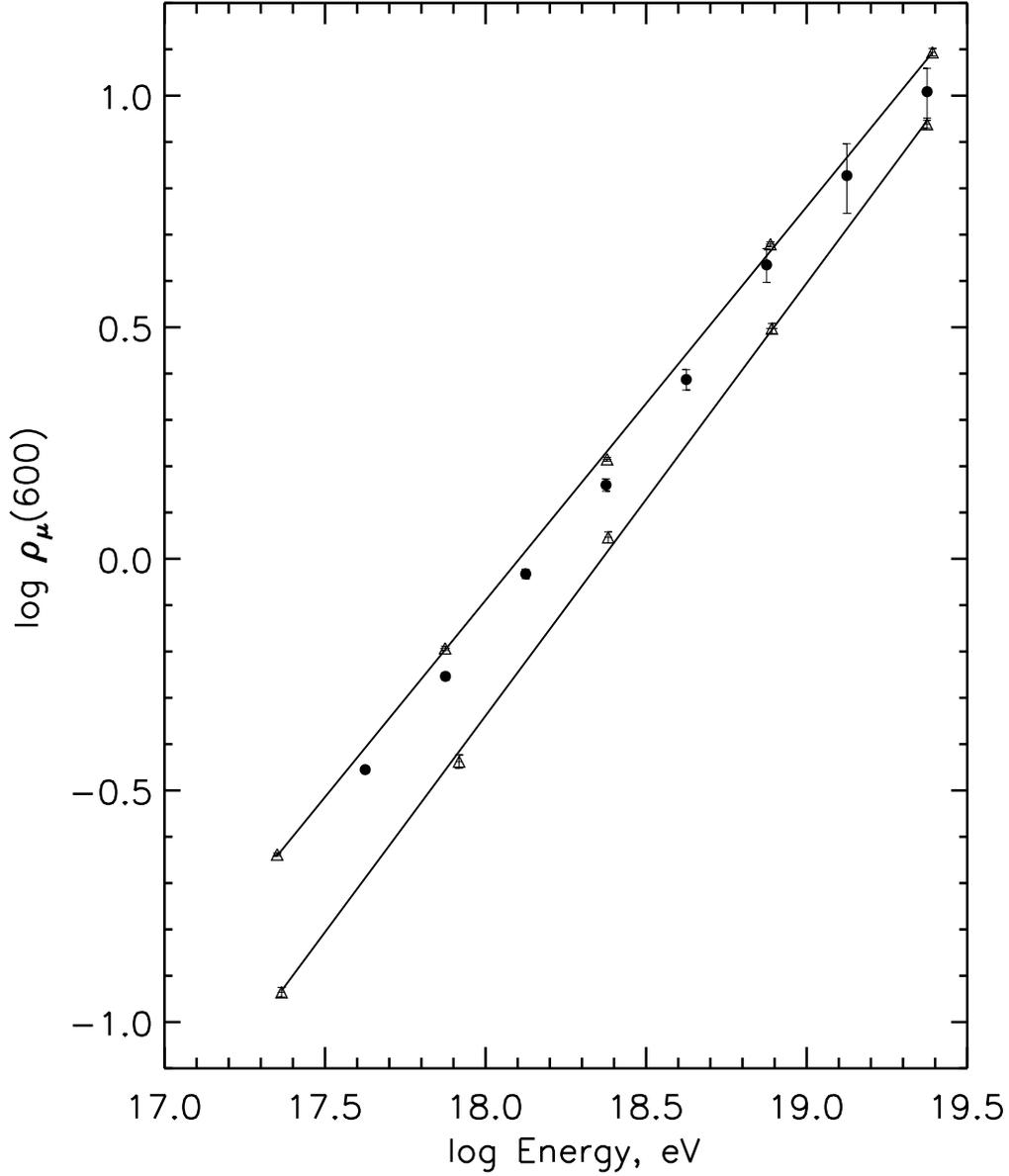,height=17cm}
\caption{Data from the AGASA A100 array (muon threshold of
0.5\,GeV) from \protect\cite{Haya97}, with the 1.4
multiplicative factor removed (see text).  Sold lines and
triangles show simulations from MOCCA+SIBYLL for iron (top) and
proton (bottom) primary particles.  A small triggering bias is present
in A100 at lower energies which cannot be indicated here.  The
simulations were performed at energies of $3\times10^{17}$eV, 
$10^{18}$eV, $3\times10^{18}$eV,$10^{19}$eV and $3\times10^{19}$eV }
\end{figure}

\begin{figure}[h]
\vspace{-1.4cm}
\epsfig{figure=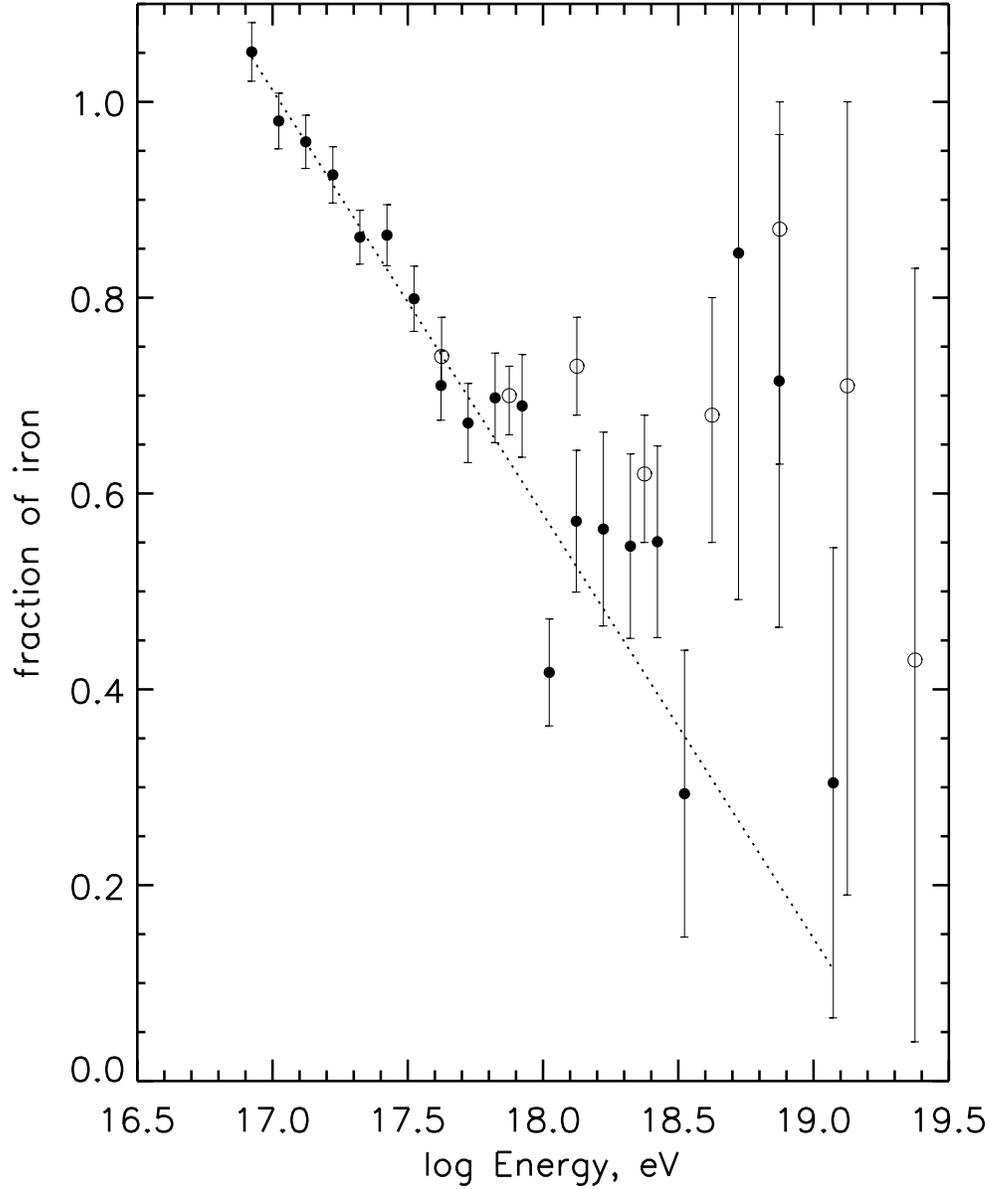,height=17cm}
\caption{Predicted fraction of iron nuclei in the cosmic ray beam
at the top of the atmosphere, from the AGASA A1 (filled circles)
and A100 (open circles) experiments.  We assume a simple
proton/iron mixture, and the hadronic physics contained in
SIBYLL.  Error bars are calculated using the uncertainties in the
experimental muon densities and uncertainties in the logarithmic
fits to the Monte Carlo predictions in Figures~2~\&~3.  The
dotted line represents a fit to the A1 points, a fit mainly
influenced by the lower energy data.  The results from the two
arrays (with different muon energy thresholds) appear consistent
in the region where they overlap.}
\end{figure}

\begin{figure}[h]
\vspace{-1.4cm}
\epsfig{figure=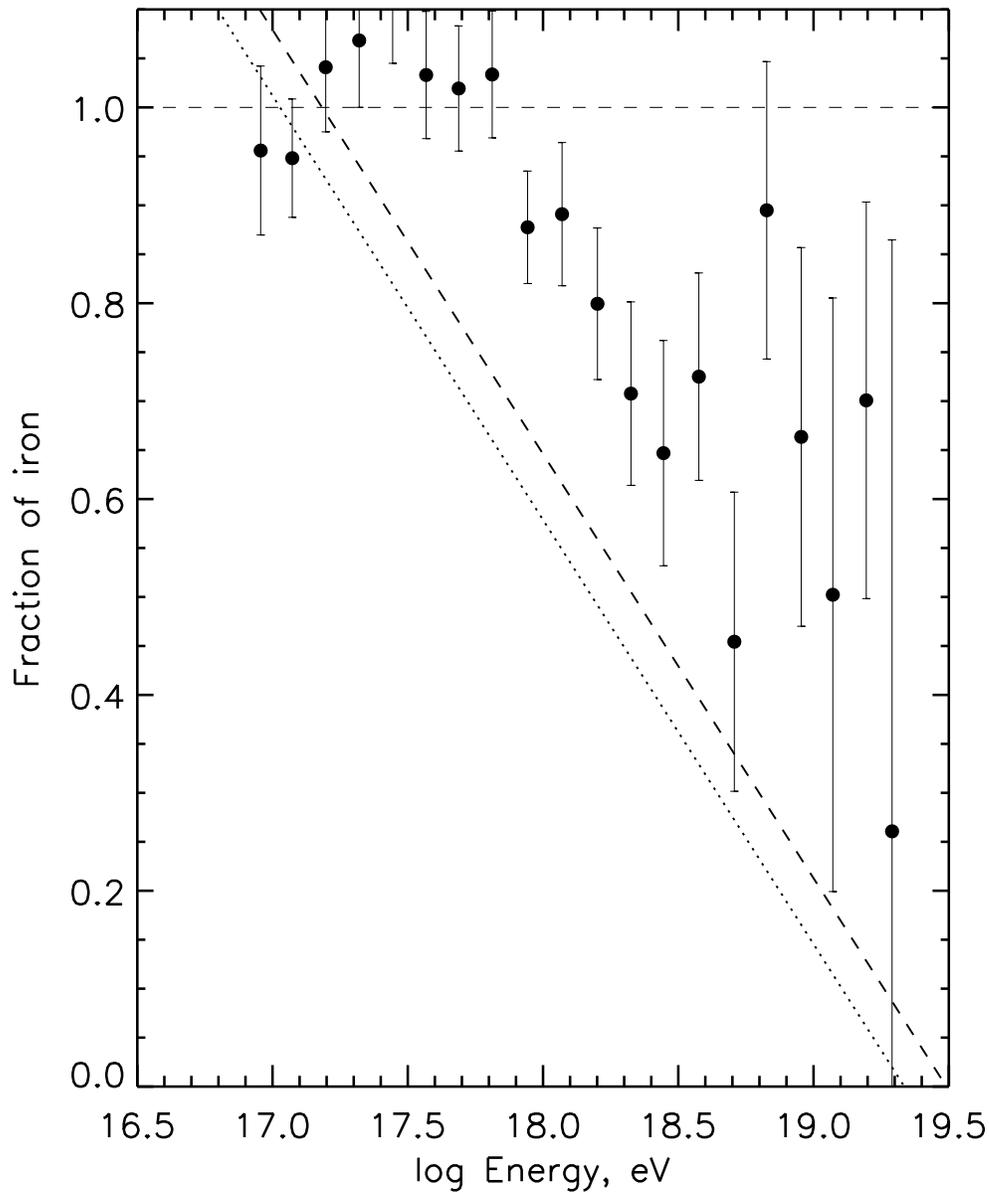,height=17cm}
\caption{Predicted fraction of iron nuclei in the cosmic ray beam
at the top of the atmosphere, from the Fly's Eye data under the
assumption of the SIBYLL hadronic model.  A simple two-component
composition model is assumed.  Errors are calculated using the data-point
error estimates, and errors from the logarithmic fits to the Monte Carlo
proton and iron expectations.  The dotted line shows a fit to the
A1 data from Figure~4.  The dashed line is that same fit shifted under the
assumption that the A1 analysis underestimates primary energy by 30\%.}
\end{figure}

\begin{figure}[h]
\vspace{-1.4cm}
\epsfig{figure=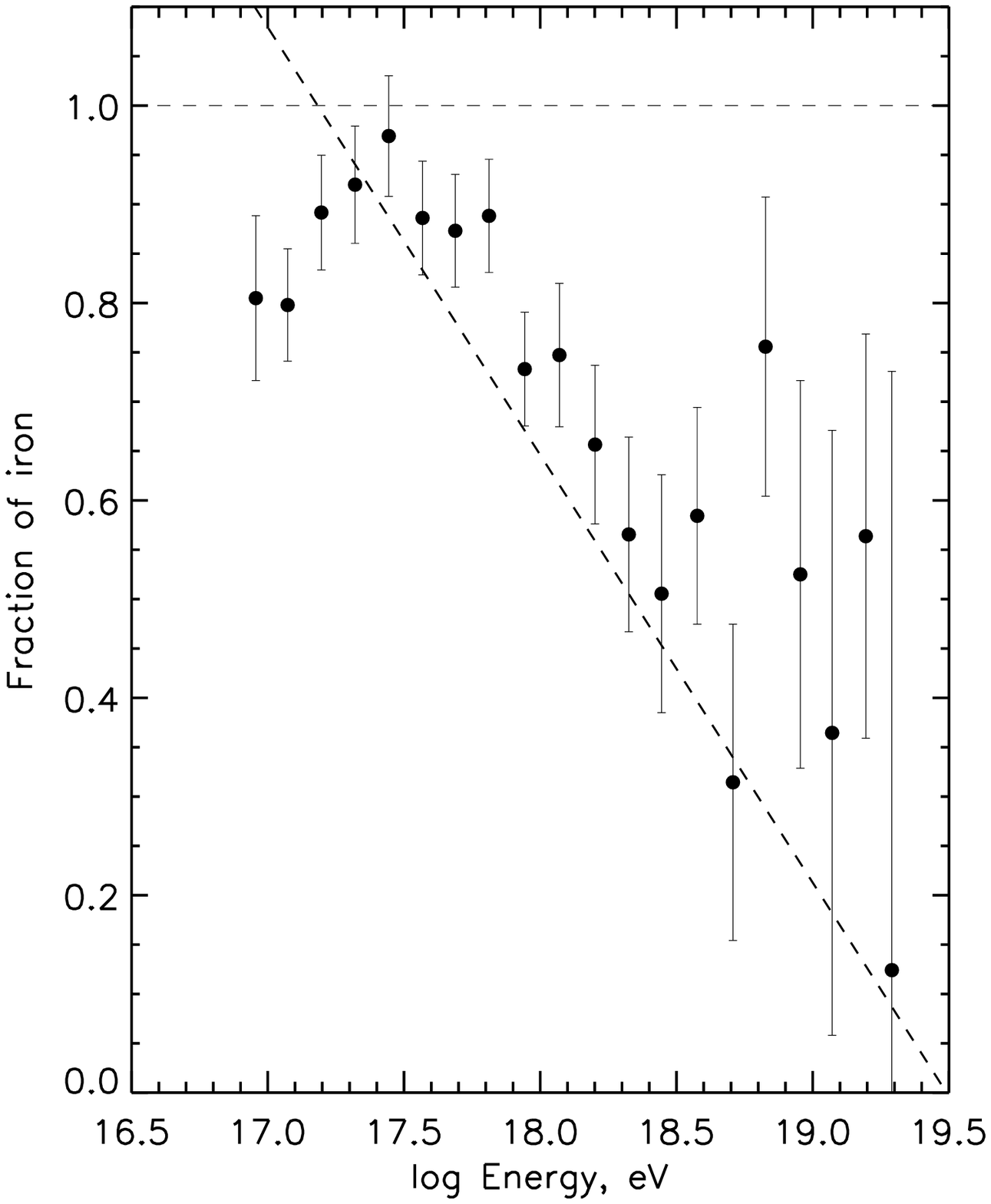,height=17cm}
\caption{Like Figure~5, but here we attempt to show the effect of
possible systematic errors in the data or hadronic model.  We
assume that the A1 analysis underestimates energy by 30\%.  We
also assume that there is a further 10\,g\,cm$^{-2}$ systematic
shift required in either the Fly's Eye data or in the $X_{\rm
max}$ simulation results.}
\end{figure}


\begin{thebibliography}{999}
\bibitem{Balt85} Baltrusaitis,\,R.M. et al., Nucl. Instrum. Methods  {\bf{A240}}, 410 (1985)
\bibitem{AGASA} Chiba,\,N. et al., Nucl. Instrum. Methods {\bf A311}, 338 (1992). 
\bibitem{Gaiss93} Gaisser,\,T. K.  et al., Physical Review {\bf D47}, 1919 (1993). 

\bibitem{Todor} Stanev,\,T., private communication.

\bibitem{Haya95} Hayashida,\,N. et al., J.\,Phys.\,G. {\bf 21}, 1101 (1995). 
\bibitem{Flet94} Fletcher,\,R. S. et al., Physical Review {\bf D50}, 5710 (1994). 
\bibitem{Hill81} Hillas,\,A. M., Proceedings of the $17^{\rm th}$ Int. Cosmic Ray Conf., Paris {\bf 11}, 193  (1981). 
\bibitem{A1} Hara,\,T. et al., Proceedings of the $16^{\rm th}$ Int. Cosmic Ray Conf., Kyoto {\bf 8}, 135  (1979). 
\bibitem{Haya97} Hayashida,\,N. et al.,  Proceedings of the $25^{\rm th}$ Int. Cosmic Ray Conf., Durban {\bf 6}, 241  (1997). 
\bibitem{1.4} Matsubara,\,Y. et al., Proceedings of the $19^{\rm th}$ Int. Cosmic Ray Conf., La Jolla {\bf 7}, 119  (1985).
\bibitem{attenn} Yoshida,\,S. et al., J.\,Phys.\,G. {\bf 20}, 651 (1994).
\bibitem{energy} Yoshida,\,S. et al., Astropart.\,Phys. {\bf 3}, 105 (1995).
\bibitem{dai88} Dai,\,H.Y. et al., J.\,Phys.\,G. {\bf 14}, 793 (1988).
\bibitem{Ding97} Ding, L. K. et al., Astrophys. J. {\bf 474}, 490 (1997).  


\end{thebibliography}
\end{document}